\documentstyle[12pt]{article}

\begin{document}
\begin{center}
{\Large \bf Anomalous effects due to the inertial
anti-gravitational potential of the sun}

\bigskip

{\large D.L.~Khokhlov}
\smallskip

{\it Sumy State University, Ukraine\\
E-mail: dlkhokhl@rambler.ru}
\end{center}

\begin{abstract}
It is introduced inertial anti-gravitational potential into
the theory of gravity to stop gravitational collapse at the
nuclear density and thus prevent singularities. It is considered
effective gravity which includes Newtonian potential and
inertial anti-gravitational potential. It is investigated
footprints of the effective gravity in the solar system. The
inertial anti-gravitational potential of the sun allows to explain
the anomalous acceleration of Pioneer 10 and 11, the anomalous
increase in the lunar semi-major axis,
the residuals of the seasonal
variation of the proper angular velocity of the earth,
the anomalous increase of the Astronomical Unit,
the anomalous shift of the perihelion of mercury.

\end{abstract}

\section{Basic idea}

The theory of gravity~\cite{mtw} faces the problem of
singularities. In particular, singularities may arise as a result
of the gravitational collapse of massive stars~\cite{Sha}.
A body contracted to the nucleus density resembles the neutron
star. There is a maximum mass for the neutron star of
order of the mass of the sun, $m_{max}\sim m_{\odot}$. For the
neutron star with the mass less than the maximum mass, the
pressure due to the degenerated neutron Fermi gas balances the
gravity of the star. For the neutron star with the mass more than
the maximum mass, the gravity of the star overcomes the pressure
due to the degenerated neutron Fermi gas, and the star goes to the
singularity.

As known the nucleus of atom consists of elementary particles,
protons and neutrons. Those have the intrinsic angular momentum,
spin
\begin{equation}
\frac{\hbar}{2}=m_{p}vr_p
\label{eq:sp}
\end{equation}
where $\hbar$ is the Planck constant, $m_p$ is the mass of proton,
$r_p$ is the radius of proton.
The spin of proton (neutron) yields the centrifugal acceleration
\begin{equation}
w_{p}=\frac{v^2}{r_p}=\frac{\hbar^2}{4m_p^2r_p^3}.
\label{eq:wp}
\end{equation}
This centrifugal acceleration opposes acceleration due to gravity.
Consider the neutron star and compare the gravity of the star and
the centrifugal acceleration due to the spin of proton (neutron).
In view of eq.~(\ref{eq:wp}) the centrifugal acceleration due to the
spin of proton is of order
$w_{p}\sim 10^{31}\ \mathrm{cm/s^2}$.
Take the maximum possible neutron star of the mass of
the visible universe. While adopting the size of
the visible universe $r_U\sim 10^{28}\ \mathrm{cm}$,
the mass of the visible universe is
$m_U=c^2r_U/G\sim 10^{56}\ \mathrm{g}$.
While adopting the density of neutron star
$\rho_{NS}\sim 10^{14}\ \mathrm{g/cm^3}$,
the radius of neutron star for the mass of the visible
universe is of order
$r_{UNS}=(3m_{U}/4\pi\rho_{NS})^{1/3}\sim 10^{13}\ \mathrm{cm}$.
The acceleration due to gravity of the neutron star of the mass of
the visible universe is of order
$w_{UNS}=Gm_{U}/r_{UNS}^2\sim 10^{21}\ \mathrm{cm/s^2}$.
Thus the centrifugal acceleration due to the spin of proton
exceeds the gravity of neutron star of any possible mass in
the visible universe. The centrifugal acceleration due to the spin
of proton prevents collapse of a neutron star.
Then one can consider proton incompressible in the gravitational
interaction.

The results of experiments show that in electromagnetic, weak,
strong interactions elementary particles do not change their size
and mass with the momentum conservation law holding true. This
means that elementary particles are incompressible in
electromagnetic, weak, strong interactions.
Let two protons collide with the velocity of light $c$.
Estimate acceleration experienced by the
proton under proton colliding. The radius of
strong interaction is of order of the size of proton,
$r_p\sim 10^{-13}$ cm. Then the acceleration experienced by the
proton is $w=c^2/r_p\sim 10^{34}\ \mathrm{cm/s^2}$.
The acceleration due to gravity of the neutron star of the mass of
the visible universe
$w_{UNS}\sim 10^{21}\ \mathrm{cm/s^2}$
is much less than that in the strong interaction under proton
colliding.

So elementary particles are incompressible hence the neutron star
consisting of elementary particles is incompressible.
Then the radius of neutron star is a limiting one under
contraction of a body with the nuclear density being a limiting
one. All the atomic nuclei of a body specifies incompressible
volume of the radius of neutron star for a body.
Incompressibility of the volume of the radius of neutron star
means that the total force at the radius of
neutron star must be equal to zero. Then
there must be an inertial force to balance the
force of gravity at the radius of neutron star.
Suppose that a body produces inertial anti-gravitational
potential
\begin{equation}
\Psi=\frac{4\pi}{3}G\rho r_{NS}^2
\label{eq:Psi}
\end{equation}
where $G$ is the Newtonian constant,
$\rho$ is the density of the body,
$r_{NS}$ is the radius of neutron star for the body.
We arrive at the effective gravity which includes
Newtonian potential and inertial anti-gravitational potential
\begin{equation}
\Phi=-\frac{Gm}{r} + \Psi.
\label{eq:phi}
\end{equation}
For the body contracted to the nuclear density the inertial
anti-gravitational potential balances the Newtonian potential thus
preventing singularities under the gravitational collapse of the
body.

Introduction of the inertial anti-gravitational potential means
that we consider the gravitating body in the non-inertial frame.
Let us start with the Lagrangian of a particle of the unit mass
moving in the field of a gravitating body. In the background frame
the Lagrangian is given by
\begin{equation}
L=\frac{v^2}{2}+\frac{Gm}{r}
\label{eq:L}
\end{equation}
One can write down inertial anti-gravitational potential as
\begin{equation}
\Psi=\frac{v_\Psi^2}{2}.
\label{eq:psiv}
\end{equation}
This means the gravitating body brings in the frame with an
effective velocity $v_\Psi$ directed from the
centre of the body along the radial coordinate.
While taking the gravitating body as a rest frame then
a particle is in the frame with an
effective velocity $v_\Psi$ directed towards the body.
The velocity of a particle with respect to the
background frame is
\begin{equation}
v=v^\prime-v_\Psi.
\label{eq:v}
\end{equation}
Substituting eq.~(\ref{eq:v}) into eq.~(\ref{eq:L}) and following
the standard procedure~\cite{Lan} we obtain the Lagrangian in the
frame with an effective velocity $v_\Psi$
\begin{equation}
L^\prime=\frac{v^{\prime\,2}}{2}+\frac{Gm}{r^\prime}+
\frac{dv_\Psi}{dt}r^\prime
\label{eq:L2}
\end{equation}
where $dv_\Psi/dt$ is the inertial acceleration directed backwards
the body. In view of eq.~(\ref{eq:phi}) the inertial acceleration
is given by
\begin{equation}
w_{in}\equiv\frac{dv_\Psi}{dt}=\frac{\Psi}{r^\prime}.
\label{eq:win}
\end{equation}
From the Lagrangian eq.~(\ref{eq:L2}) one can derive the equation
of motion
\begin{equation}
w=-\frac{Gm}{r^{\prime\,2}}+ \frac{\Psi}{r^\prime}
\label{eq:w}
\end{equation}
where the first term is the Newtonian acceleration due to gravity,
the second term is the inertial acceleration due to the inertial
anti-gravitational potential.
Thus introduction of the inertial anti-gravitational potential
does not modify Newtonian gravity.

Let the sun along with the Newtonian potential produce the
inertial anti-gravitational potential. Estimate the inertial
anti-gravitational potential of the sun. Calculate the density of
neutron star as
$\rho_{NS}=3m_p/4\pi r_0^3=3.0\times 10^{14}\ \mathrm{g/cm^3}$
where $m_p$ is the mass of proton,
$r_0=1.1\times 10^{-13}\ \mathrm{cm}$ is the radius of the
nucleus. Then the radius of neutron star for the sun is equal to
$r_{NS}=(3m_{\odot}/4\pi\rho_{NS})^{1/3}=
1.2\times 10^{6}\ \mathrm{cm}$.
In view of eq.~(\ref{eq:Psi}), the inertial anti-gravitational
potential of the sun is equal to
$\Psi_{\odot}=5.3\times 10^{5}\ \mathrm{cm^2/s^2}$.

\section{Anomalous acceleration of Pioneer 10 and 11}

The inertial acceleration due to the inertial anti-gravitational
potential gives contribution into the first order relativistic
effect. The inertial acceleration of the earth due to
the inertial anti-gravitational potential of the sun is given by
$w_E=\Psi_{\odot}/r_{SE}$ where $r_{SE}$ is the distance
between the earth and sun.
This acceleration can be seen in ranging to distant spacecraft
as a blue shift of the frequency of the electromagnetic field
\begin{equation}
\frac{\Delta\omega}{\omega}\approx\frac{w_Et}{c}=
\frac{\Psi_{\odot}t}{cr_{SE}}.
\label{eq:we}
\end{equation}
In ranging to distant spacecraft, the acceleration
of the earth outward the sun looks like the acceleration of the
spacecraft inward the sun, $w_{sc}=2w_E$. The factor 2
takes into account that the acceleration of the
earth gives contribution into the shift of the reference frequency
during the time of two-leg light travel while the acceleration of
the spacecraft gives contribution into the shift of the observed
re-transmitted frequency during the time of one-leg light travel.

Analysis of radio Doppler and ranging data from distant
spacecraft~\cite{An} indicated that an
anomalous inward acceleration is acting on Pioneer 10 and 11,
$w_P=(8.74\pm 1.25)\times 10^{-8}\ \mathrm{cm/s^2}$.
This may be interpreted as the outward acceleration of the earth.
In view of eq.~(\ref{eq:we}), determine the inertial
anti-gravitational potential of the sun from the anomalous
acceleration of Pioneer 10 and 11. Then we obtain
$\Psi_{\odot}=2w_Pr_{SE}=6.6\times 10^5\ \mathrm{cm^2/s^2}$.

\section{Polarization of the moon's orbit}

The potential $\Psi_{\odot}$ gives contribution into the
polarization of the moon's orbit in the direction of the sun.
While adopting the sum, $Gm_{\odot}/r_{SE}+\Psi_{\odot}$, as an
effective
gravitational potential of the sun at the radius $r_{SE}$, we
reveal an additional inertial potential along the moon's orbit
\begin{equation}
V\approx\frac{\Psi_{\odot}r_{EM}}{r_{SE}}\sin\Omega_{ME} t
\label{eq:V}
\end{equation}
where $r_{EM}$ is the distance between the earth and moon,
$\Omega_{ME}$ is the angular velocity of the moon.
An observer at the earth treats this as an additional outward
acceleration of the moon
\begin{equation}
w_M\approx\frac{\Psi_{\odot}}{r_{SE}}|\sin\Omega_{ME} t|
\label{eq:dwm}
\end{equation}
where $|\sin|$ is the modulus of the function.
The additional outward acceleration of the moon average for the
period of revolution of the moon is given by
\begin{equation}
w_{M}^{av}\approx\frac{\Psi_{\odot}}{\sqrt{2}r_{SE}}.
\label{eq:adwm}
\end{equation}
This can be seen in lunar laser ranging as a velocity
\begin{equation}
v_{M}^{av}=\frac{w_{M}^{av}r_{EM}}{c}=
\frac{\Psi_{\odot}r_{EM}}{\sqrt{2}cr_{SE}}.
\label{eq:adv}
\end{equation}
There is a difference in the rate of the lunar semi-major axis
increases obtained from telescopic observations and from lunar
laser ranging,
$\dot{a}_{LLR}-\dot{a}_{tel}=1.29\ \mathrm{cm/yr}=
4.1\times 10^{-8}\ \mathrm{cm/s}$~\cite{Dum}.
It is used telescopic observations of the secular deceleration of
the proper angular velocity of the earth from which the rate of
the lunar semi-major axis increase is determined.
In view of eq.~(\ref{eq:adv}), determine the inertial
anti-gravitational potential of
the sun from this value. Then we obtain
$\Psi_{\odot}=6.9\times 10^5\ \mathrm{cm^2/s^2}$.

\section{Seasonal variation of the angular velocity of the
earth}

Under the variation of the earth-sun distance the inertial
anti-gravitational potential
of the sun yields the variation of the
angular velocity of the earth around the sun
\begin{equation}
\frac{\Delta \Omega_{ES}}{\Omega_{ES}}=-
\frac{\Psi_{\odot}r_{SE}}{Gm_\odot}e\cos\varphi
\label{eq:Dom}
\end{equation}
where $e$ is the eccentricity, $\varphi$ is the angle (longitude)
of the earth's orbit around the sun.
This looks like the variation of the proper
angular velocity of the earth with respect to remote stars
\begin{equation}
\frac{\Delta \Omega_{E}}{\Omega_{E}}\approx -
\frac{\Psi_{\odot}r_{SE}}{Gm_\odot}e\cos 23.5^0\cos\varphi
\label{eq:Dom2}
\end{equation}
where $\cos 23.5^0$ accounts for the tilt of the earth's axis of
rotation to the perpendicular to the earth-sun plain.
Observation~\cite{DF} gives the residuals of the seasonal
variation of the proper angular velocity of the earth,
$\Delta \Omega_{E}/\Omega_{E}=
-(9\pm 2)\times 10^{-10}\cos\varphi$.
In view of eq.~(\ref{eq:Dom2}), determine the inertial
anti-gravitational potential of the sun from this value.
Then we obtain
$\Psi_{\odot}=5.2\times 10^5\ \mathrm{cm^2/s^2}$.

\section{Secular increase of Astronomical Unit}

Under the variation of the earth-sun distance the inertial
anti-gravitational potential of the sun is seen as an anomalous
potential of the earth
\begin{equation}
V_E=\Psi_{\odot}e|\cos\varphi|.
\label{eq:dve}
\end{equation}
This gives contribution into a second order relativistic shift of
the reference frequency in the planet ranging
\begin{equation}
\frac{\Delta\omega}{\omega}=\frac{\Psi_{\odot}e}{\sqrt{2}c^2}
\label{eq:psie}
\end{equation}
where the factor $1/\sqrt{2}$ accounts for averaging for
the period. While interpreting the shift as a first order
effect this mimics the velocity of the earth outward the sun
\begin{equation}
v_E=\frac{\Psi_{\odot}e}{\sqrt{2}c}.
\label{eq:dau}
\end{equation}

Krasinsky and Brumberg reported~\cite{KB} the anomalous positive
secular trend in the Astronomical Unit
$(d/dt)AU=15\pm 4\ \mathrm{m/cy}$ obtained from analysis of the
data on ranging to the major planets (mars, venus, mercury).
Pitjeva gets the smaller value about 5 metres per century, see
discussion to the talk by Standish~\cite{St}.
In view of eq.~(\ref{eq:dau}), determine the velocity of the earth
adopting the inertial anti-gravitational potential of
the sun $\Psi_{\odot}=6.6\times 10^5\ \mathrm{cm^2/s^2}$.
Then we obtain
$v_E=2.6\times 10^{-7}\ \mathrm{cm/s}=8.3\ \mathrm{m/cy}$
that may explain the anomalous secular trend in the Astronomical
Unit.

\section{Advance of the perihelion of mercury}

The inertial anti-gravitational potential of the sun should lead
to the precession of the Keplerian orbit of a planet under its
motion in the gravitational field of the sun. A small addition
$\Psi_{\odot}$ to
the potential of the sun causes the shift of the perihelion of
planet's orbit per revolution by the value~\cite{Lan}
\begin{equation}
\delta\varphi=\frac{\partial}{\partial M}\frac{2m^2}
{M}\int\limits_{0}^{\pi}r^2 \Psi_{\odot} d\varphi
\label{eq:dphi}
\end{equation}
where $m$ is the mass of the planet, $M$ is the angular
momentum.
The non-perturbed orbit is given by
\begin{equation}
r=\frac{p}{1+e\cos\varphi}
\label{eq:orb}
\end{equation}
with
\begin{equation}
p=\frac{M^2}{Gm^2m_{\odot}} \qquad
p=a(1-e^2)
\label{eq:p}
\end{equation}
where $p$ is the orbit's latus rectum,
$a$ is the semi-major axis.
Integration of eq.~(\ref{eq:dphi}) with the use of
eqs.~(\ref{eq:orb}),(\ref{eq:p})
gives the shift (advance) of the perihelion of a planet due to
the potential $\Psi_{\odot}$ per revolution
\begin{equation}
\delta\varphi\approx\frac{6\pi a(1-e^2)\Psi_{\odot}}{Gm_{\odot}}.
\label{eq:dphi2}
\end{equation}
Determine the inertial anti-gravitational potential of
the sun from the data on the anomalous shift of the perihelion of
mercury, $43$ arcseconds~\cite{mtw}. Then we obtain
$\Psi_{\odot}=6.4\times 10^5\ \mathrm{cm^2/s^2}$.

\section{Discussion}

We have considered effective gravity which includes Newtonian
potential and inertial anti-gravitational potential. We have
investigated footprints of the inertial anti-gravitational
potential of the sun. We have shown that the inertial
anti-gravitational potential of the sun allows to explain
the anomalous acceleration of Pioneer 10 and 11,
the anomalous increase in the lunar semi-major axis,
the residuals of the seasonal
variation of the proper angular velocity of the earth,
the anomalous increase of the Astronomical Unit,
the anomalous shift of the perihelion of mercury.
The theoretical estimate of the inertial
anti-gravitational potential of the sun is
$\Psi_{\odot}=5.3\times 10^{5}\ \mathrm{cm^2/s^2}$.
The inertial anti-gravitational potential of the sun determined
from the anomalous acceleration of Pioneer 10 and 11 is
$\Psi_{\odot}=6.6\times 10^5\ \mathrm{cm^2/s^2}$,
from the anomalous increase in the lunar semi-major axis
$\Psi_{\odot}=6.9\times 10^5\ \mathrm{cm^2/s^2}$,
from the residuals of the seasonal
variation of the proper angular velocity of the earth
$\Psi_{\odot}=5.2\times 10^5\ \mathrm{cm^2/s^2}$,
from the anomalous shift of the perihelion of mercury
$\Psi_{\odot}=6.4\times 10^5\ \mathrm{cm^2/s^2}$.
The inertial anti-gravitational potential of the sun is seen
as an anomalous potential of the earth in the planet ranging.
This mimics the velocity of the earth outward the sun and may
explain the anomalous increase of the Astronomical Unit.
Thus the data from the above observations may be considered as a
support for the effective gravity with the inertial
anti-gravitational potential.

The Shapiro effect~\cite{mtw} contributing to ranging is a
powerful tool for exploring gravity. The Shapiro effect being
dependent of the gravitational potential is not sensible to the
inertial anti-gravitational potential because it is fixed with
radius. Thus the inertial anti-gravitational potential cannot be
seen in ranging data. It is worth noting that the authors
of~\cite{An} consider the phenomenological model, p.11.5
quadratic in time model, which fits well the Pioneer effect
and ranging data. This model adds a quadratic in time term to the
light time as seen by the DSN station. It mimics a line of sight
acceleration of the spacecraft and could be thought of as an
expanding space model. Phenomenologically this
model is similar to the effective gravity with the inertial
anti-gravitational potential. The expansion of space may be
interpreted as an effective velocity due to the
inertial anti-gravitational potential.

The inertial anti-gravitational potential of the sun provides an
explanation of the anomalous shift (advance) of the perihelion of
mercury alternative to the general relativity~\cite{mtw}.
The theory of relativity may be reinterpreted in a way
that relativistic effects pertain only to electromagnetism while
gravitation remains non-relativistic. In this case
effects of general relativity such as
gravitational redshift, time delay due to gravity arise due to
extension of special relativity with the use of the principle of
equivalence.
The motion of bodies in the gravitational potential is
described within the Newtonian mechanics.

\end{document}